\documentclass[prl,letterpaper,twocolumn,showpacs,superscriptaddress,amsmath,amsfonts,amssymb,preprintnumbers,citeautoscript]{revtex4}
\pdfoutput=1
\usepackage[T1]{fontenc}\usepackage[latin1]{inputenc}
\usepackage{dcolumn,graphicx,color,booktabs,microtype,afterpage} 
\usepackage[charter]{mathdesign}
\renewcommand{\figurename}{Fig.}
\renewcommand{\tablename}{Table}
\makeatletter\renewcommand{\fnum@figure}[1]{\figurename~\thefigure.}\makeatother
\makeatletter\renewcommand{\fnum@table}[1]{\tablename~\thetable.}\makeatother
\definecolor{MyRed}{rgb}{0.8,0,0}
\definecolor{MyGreen}{rgb}{0.25,0.5,0.5}

\usepackage[colorlinks,plainpages=false,linkcolor=blue,urlcolor=blue,citecolor=black,pdfpagemode=UseNone,pdfstartview=FitBH]{hyperref}

\begin{document} \pagestyle{plain}

\title{Excitation energy map of the high-energy dispersion anomalies in cuprates.}

\author{D.\,S.\,Inosov}\author{R.\,Schuster}
\affiliation{Institute for Solid State Research, IFW Dresden, P.\,O.\,Box 270116, D-01171 Dresden, Germany.}
\author{A.\,A.\,Kordyuk}
\affiliation{Institute for Solid State Research, IFW Dresden, P.\,O.\,Box 270116, D-01171 Dresden, Germany.}
\affiliation{Institute of Metal Physics of National Academy of Sciences of Ukraine, 03142 Kyiv, Ukraine.}
\author{J.\,Fink}
\affiliation{BESSY GmbH, Albert-Einstein-Strasse 15, 12489 Berlin, Germany.}%
\affiliation{Institute for Solid State Research, IFW Dresden, P.\,O.\,Box 270116, D-01171 Dresden, Germany.}
\author{S.\,V.~Borisenko}\author{V.\,B.\,Zabolotnyy}\author{D.\,V.~Evtushinsky}\author{M.\,Knupfer}\author{B.\,Büchner}
\affiliation{Institute for Solid State Research, IFW Dresden, P.\,O.\,Box 270116, D-01171 Dresden, Germany.}
\author{R.\,Follath}
\affiliation{BESSY GmbH, Albert-Einstein-Strasse 15, 12489 Berlin, Germany.}
\author{H.\,Berger}
\affiliation{Institut de Physique de la Mati\`{e}re Complexe, EPFL, 1015 Lausanne, Switzerland.}




\begin{abstract}
The anomalous high-energy dispersion of the conductance band in the high-$T_\text{c}$ superconductor
Bi(Pb)$_2$Sr$_2$CaCu$_2$O$_{8+\delta}$ (Pb-Bi2212) has been extensively mapped by angle-resolved photoemission (ARPES)
as a function of excitation energy in the range from 34 to 116~eV. Two distinctive types of dispersion behavior are
observed around 0.6~eV binding energy, which alternate as a function of photon energy. The continuous transitions
observed between the two kinds of behavior near 50, 70, and 90~eV photon energies allow to exclude the possibility that
they originate from the interplay between the bonding and antibonding bands. The effects of three-dimensionality can
also be excluded as a possible origin of the excitation energy dependence, as the large period of the alterations is
inconsistent with the lattice constant in this material. We therefore confirm that the strong photon energy dependence
of the high-energy dispersion in cuprates originates mainly from the photoemission matrix element that suppresses the
photocurrent in the center of the Brillouin zone.
\end{abstract}


\maketitle

The anomalous high-energy dispersion in the electronic structure of cuprates remains a hot topic in the high-temperature
superconductivity research \nocite{Inosov07, VallaKidd06, GrafMcElroy06, XieYangShen06, PanRichard06, GrafLanzara07,
MeevasanaZhou06, ChangPailhes06, RonningShen05, HwangNicol07, ByczukKollar07, Macridin07, Srivastava07, ZhouWang07,
Markiewicz07, ZhuAji08, CojocaruCitro07, LeighPhillips07, TanWan07, Manousakis07, AlexandrovReynolds07,
MarkiewiczBansil07, ZhangLiu08, MeevasanaBaumberger08}[1--24]. After multiple attempts to explain this phenomenon as an
intrinsic property of the spectral function, it was finally shown that the experimentally observed dispersion
significantly depends on the experimental conditions, such as photon energy and the experimental geometry
\cite{Inosov07, ZhangLiu08}, which suggested that the influence of photoemission matrix elements distorts the real
behavior of the conductance band at high binding energies. It is therefore essential to gain a deeper understanding of
this effect in order to uncover the underlying electronic structure.

Two seemingly reasonable explanations for these chan\-ges in behavior could be related to \cite{MeevasanaBaumberger08}
(i) bilayer splitting, i.e. modulation of the relative intensity of the bonding and antibonding bands due to the
photoemission matrix elements; (ii) effects of the $k_z$ dispersion that cause periodic changes of the ARPES signal with
varying excitation energy. In the following, we will show that both these hypotheses are inconsistent with the
experimental observations.

\begin{figure}[b]
\includegraphics[width=\columnwidth]{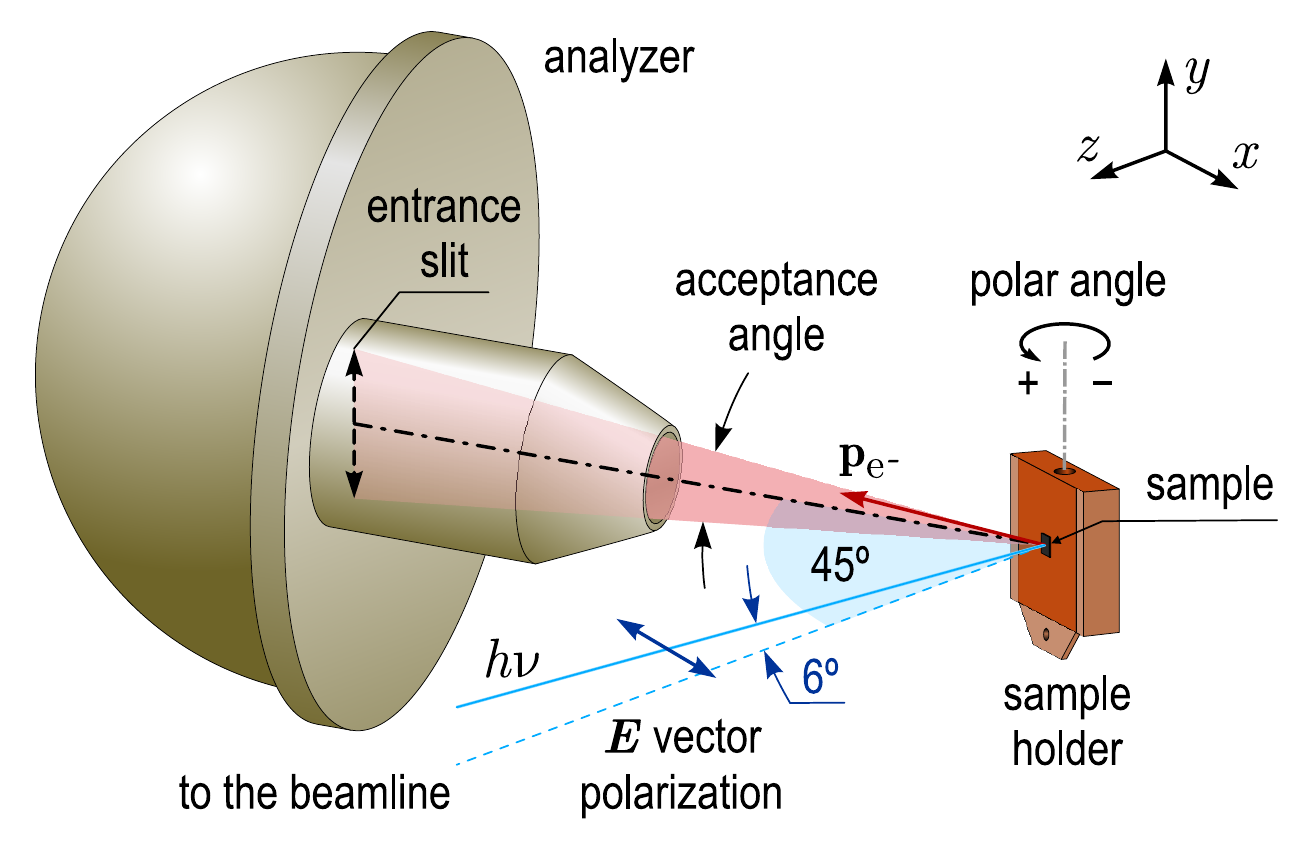} \caption{Sketch of the experimental geometry.} \label{Fig:Geometry}
\end{figure}

\begin{figure*}[t]
\includegraphics[width=0.7\textwidth]{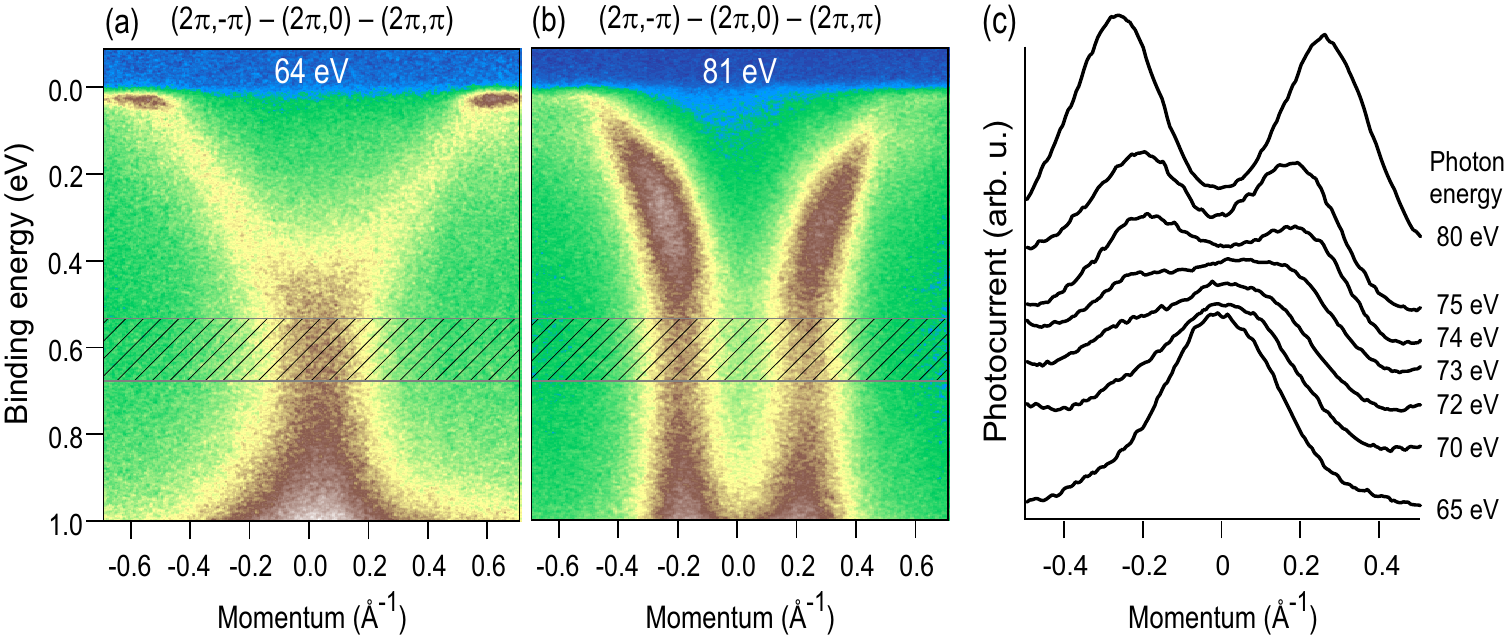} \caption{Photon energy dependence of the high-energy
anomaly in Pb-Bi2212. A pair of equivalent spectra taken in the 2nd BZ along the
$(2\pi,-\pi)$\,--\,$(2\pi,0)$\,--\,$(2\pi,\pi)$ direction with excitation energies 64 and 81~eV are shown in panels (a)
and (b) respectively. The spectrum (a) is an example of the ``champagne glass'' dispersion, while spectrum (b)
represents the ``waterfalls'' behavior. The momentum distribution curves integrated in a small energy window around
0.6~eV binding energy (hatched area) are shown in panel (c) for a number of excitation energies, showing a transition
between the two types of behavior at about 70~eV.} \label{Fig:Curves}
\end{figure*}

In this paper, we take a closer look at the excitation energy dependence of the high-energy anomaly. As was previously
reported in Ref.\,\onlinecite{Inosov07}, there are two distinctive types of behavior observed near the $\Gamma$ point in
the second Brillouin zone (BZ) in the binding energy range between 0.4 and 0.8~eV using the experimental geometry
presented in Fig.\,\ref{Fig:Geometry}. In our ARPES experiment, the optical axis of the \textit{Scienta} analyzer was
positioned at a 45$^\circ$ angle to the horizontal projection of the synchrotron beam, the beam itself was tilted by
6$^\circ$ up out of the horizontal plane, and the polarization of the photons' $\mathbf{E}$ vector was horizontal,
orthogonal to the vertical entrance slit of the analyzer. The measurements were done in the second Brillouin zone, so
that the sample was rotated from the normal emission position by a positive polar angle (towards the synchrotron beam).

Fig.\,\ref{Fig:Curves} gives an example of two equivalent ARPES spectra of slightly overdoped Pb-Bi2212
($T_\text{c}=71$\,K) taken along the $(2\pi,-\pi)$\,--\,$(2\pi,0)$\,--\,$(2\pi,\pi)$ direction in the momentum space at
two different excitation energies: 64~eV (a) and 81~eV (b). The first image shows a ``champagne glass'' type of
dispersion with a single vertical stem in the high energy region, while the second image exhibits the ``waterfalls''
behavior with two vertically dispersing features in the same energy range. In panel (c), the momentum distribution
curves (MDC) of the photocurrent integrated in a small binding energy window around 0.6~eV are plotted for several
excitation energies, showing a smooth crossover between the two types of spectra at about 72~eV. It is remarkable that
such behavior is universal for different families of cuprates \cite{Inosov07}.

\begin{figure}[b]
\hspace{-2.3pt}\includegraphics[width=1.01\columnwidth]{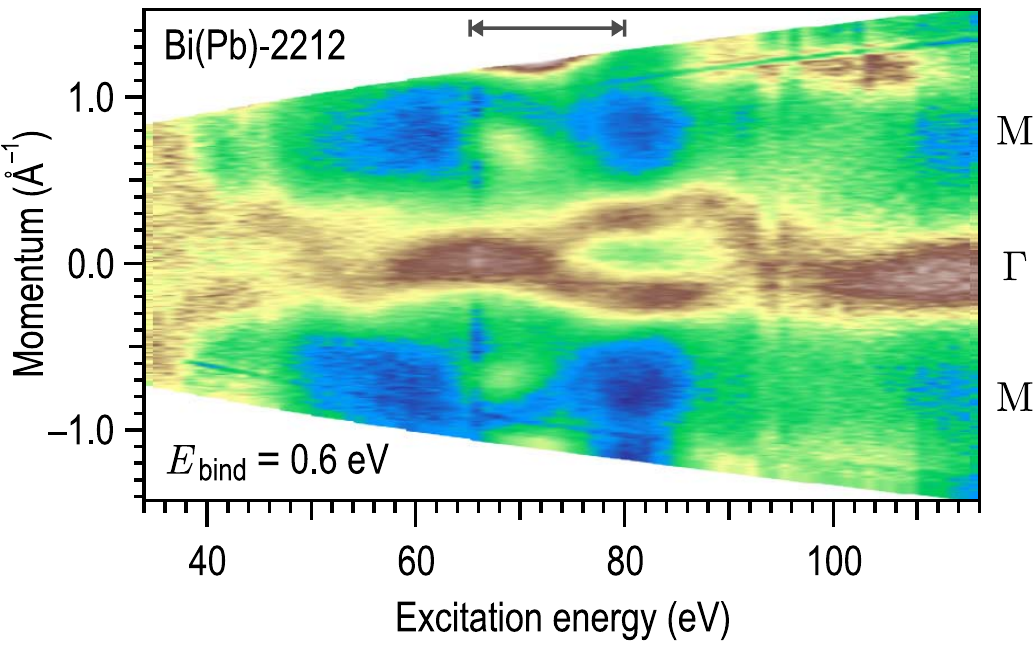} \caption{Momentum distribution of the photocurrent
along the M--$\Gamma$--M direction measured in the second Brillouin zone as a function of excitation energy, showing
several alterations of the high energy dispersion behavior. The color scale represents photoemission intensity
integrated in a small binding energy window around 0.6~eV (hatched area in Fig.~\ref{Fig:Curves}) and normalized by the
average intensity along each cut. The double-headed arrow marks the energy range covered by Fig.\,\ref{Fig:Curves}~(c).}
\label{Fig:Edep}
\end{figure}

Fig.\,\ref{Fig:Edep} shows an excitation energy map measured at 22\,K along the same cut
$(2\pi,-\pi)$\,--\,$(2\pi,0)$\,--\,$(2\pi,\pi)$ in momentum space. The color scale represents photoemission intensity
integrated in a small binding energy window around 0.6~eV. Each vertical cut corresponds to an MDC similar to those
shown in Fig.\,\ref{Fig:Curves} (c), measured with a 1~eV step in excitation energy (plotted on the horizontal axis).
The intensity of each MDC is normalized by its average value. A single MDC maximum at the $\Gamma$ point corresponds to
the ``champagne glass'' behavior, while the two split maxima represent the ``waterfalls''. Except for the already known
transition at $\sim$\,70~eV, there are two more transitions observed around 50 and 90~eV.

One can see that the distance between the MDC maxima changes continuously within each transition. The two maxima in the
``waterfalls'' region do not lose intensity, giving place to the central peak, as one would possibly expect in the case
of bilayer splitting; they rather change their position in momentum gradually, merging into a single peak. This lets us
rule out the bilayer splitting hypothesis.

It is also illustrative to compare Fig.\,\ref{Fig:Edep} to the experimentally measured photon energy dependence curves
for the matrix elements of the bonding and antibonding bands (e.g. Ref.\,\onlinecite{Kordyuk02}, Fig.\,3). The relative
intensity of the bonding band near the Fermi level is known to reach maxima at 38 and 56~eV, while the antibonding band
is enhanced by 50~eV photons. On the other hand, the transitions seen in Fig.\,\ref{Fig:Edep} do not follow this
pattern. Comparison to the theoretical photoemission intensity curves available for the bonding and antibonding bands in
an even wider photon energy window \cite{LeeFujimori02, Bansil04} will lead us to the same conclusion.

Let us now turn to the consideration of the possible role of the $k_z$-dispersion. It is well known that by varying the
excitation energy in a photoemission experiment, one can probe different $k_z$ points \cite{Huefner96}. As the Bi2212
crystals are known to be not perfectly two-dimensional \cite{Markiewicz05, Lindroos06}, this might lead to periodic
variations of the observed electronic structure as a function of photon energy. The easiest way to estimate the period
of such variations is to use the three-step model in the free electron approximation \cite{Huefner96}. The kinetic
energy of the photoelectron is given by
\begin{equation}
E_\text{kin}=(p_\perp^2+p_\parallel^2)/2m=h\nu-E_\text{bind}-\Phi\text{,}
\end{equation}
where $p_\perp$ and $p_\parallel$ are the normal and parallel components of the electron's momentum in vacuum, $h\nu$ is
the photon energy, $E_\text{bind}$ is the binding energy of the electron in the solid, and $\Phi$ is the work function.
The component of the wave vector perpendicular to the surface is given by
\begin{multline}\label{Eq:Kperp}
k_\perp\!+n_\perp\!G_\perp\!=\sqrt{\frac{2m}{\hslash^2}\,(E_\text{kin}+V_0)-(k_\parallel+n_\parallel G_\parallel)^2}\\
=\sqrt{0.262\,\frac{\,\text{\AA}^{-2\kern-5pt}}{\text{eV}}\,\,(h\nu-E_\text{bind}+V_0-\Phi)-(k_\parallel+n_\parallel
G_\parallel)^2}\text{,}
\end{multline}
where $V_0>0$ is the inner potential of the crystal, $G_\parallel$ is the reciprocal lattice vector;
$n_\perp,\,n_\parallel\in\mathbb{Z}$. At the $\Gamma$ point, $k_\parallel=0$. The periodicity in $k_\perp$ should
correspond to $G_\perp=2\pi/c=2\pi/30.89\text{\AA}\approx0.2\,\text{\AA}^{-1}$, where $c$ is the lattice constant along
$z$ direction. If the periodic changes in Fig.\,\ref{Fig:Edep} originated from the $k_z$ dispersion, one period in
$k_\perp$ would fit approximately between $h\nu_1=50$\,eV and $h\nu_2=90$\,eV. Using formula (\ref{Eq:Kperp}), we find
that this is not possible to achieve for any reasonable value of $V_0-\Phi$. Indeed, solving the equation
$k_\perp(h\nu_2)-k_\perp(h\nu_1)=0.2\,\text{\AA}^{-1}$ yields an unphysically large minimal value of $V_0-\Phi=2550$\,eV
that corresponds to $n_\parallel=0$, which lets us also reject the $k_z$ dispersion as a possible reason for the
observed changes in behavior.

We can therefore conclude that the observed photon energy dependence is most probably a photoemission matrix element
effect that suppresses the total photoemission signal near the $\Gamma$ point at particular excitation energies. This
would mean that the real underlying electronic structure is somewhere in between the ``waterfalls'' and ``champagne
glass'' types, having more spectral weight at the $\Gamma$ point than was originally observed. This is in line with the
recent result of W.~Meevasana \textit{et al.} \cite{MeevasanaBaumberger08}, who have clearly demonstrated that the
matrix element has a minimum at the center of the Brillouin zone that suppresses the spectral weight at the $\Gamma$
point. Further theoretical work needs to be done in order to understand all the details of the high-energy anomaly
behavior as a function of photon energy and gain more insight into the underlying electronic structure.

\textit{Acknowledgements}. This~project~is~part~of~the~Forschergruppe~FOR538 and is supported by the DFG under Grants
No. KN393/4. The work in Lausanne was supported by the Swiss National Science Foundation and by the MaNEP. ARPES
experiments were performed using the $1^3$ ARPES end station at the UE112-lowE PGMa beamline of the Berliner
Elektronenspeicherring-Gesellschaft für Synchrotron Strahlung m.b.H. (BESSY). We thank M.\,Lindroos for discussions and
R.\,Hübel for technical support.

\end{document}